\documentclass[10pt,a4paper]{article}
\usepackage{graphicx}
\usepackage{graphics}
\title{Lorentz Beams}
\author{Omar El Gawhary\\
Dipartimento di Fisica and Istituto Nazionale per\\ la
Fisica della Materia,
Università ``Roma Tre''\\
Via della Vasca Navale 84, I-00146 Rome, Italy \and \\
Sergio Severini \\
Centro Interforze Studi per le Applicazioni Militari\\
Via della bigattiera 10, 56010 San Piero a Grado (Pi), Italy}

\date{}
\begin{document}
\maketitle
\date{}

\begin{abstract}
A new kind of tridimensional scalar optical beams is introduced. These beams are called Lorentz beams because the form of their transverse
pattern in the source plane is the product of two independent Lorentz functions. Closed-form expression
of free-space propagation under paraxial limit is derived and pseudo non-diffracting features pointed out.
Moreover, as the slowly varying part of these fields fulfils the scalar paraxial wave equation, it follows that there exist also Lorentz-Gauss beams, i.e. beams obtained
by multipying the original Lorentz beam to a Gaussian apodization function. Although the existence of Lorentz-Gauss
beams can be shown by using two different and independent ways obtained recently from Kiselev [Opt. Spectr. {\bf 96}, 4 (2004)] and
Gutierrez-Vega {\it et al.} [JOSA A {\bf 22}, 289-298, (2005)], here we have followed a third different approach, which makes use of Lie's group theory, and which possesses the merit to put
into evidence the symmetries present in paraxial Optics.

\end{abstract}


\maketitle

\section{Introduction}
Optical beams are electromagnetic fields with a well distinguishable mean direction of propagation
 (that, from now on, we indicate as $z$ axis), in the nearness of which the most
part of field's energy is contained during propagation. After the
publication of the fundamental work due to Durnin {\it et al.}
\cite{Durnin87}, issued in 1987 about non-diffracting beams in
free-space, there was a certain number of scientifical
investigations on optical fields with the structure of beams and
which possess, or approximate at least, the diffraction features
of the aforesaid beams \cite{Vega00}-\cite{Bandres04}. Indeed, as
an ideal non-diffracting beam is physically unrealizable, because
on each plane $z=constant$ it carries an infinitive amount of
energy, it is possible to obtain diffraction-free like fields
 only with approximation: these fields are also known as pseudo non-diffracting beams. Examples of that
 are the well-known Gaussian beams, Bessel-Gauss beams \cite{Gori87} or other optical fields recently introduced
 by casting the propagation problem in coordinate systems different of rectangular and circular ones \cite{Vega05} that
 share the properties of mantaining a non-diffracting behaviour only inside a limited spatial range named Rayleigh distance.
 In the present work, we wish to introduce
 another class of pseudo non-diffracting realizable beams that we will call Lorentz beams ($LB$ for short) as well as
 their Gaussian apodizated version, that, to authors's knowledge, were never been study before today.
 If, from a theoretical point of view, the research of new kinds of optical beams is interesting, it is particularly stimulating in the present case
because of the physical realizability of the proposed field. This
realizability is not due to a Gaussian term, as
  usually happen for almost all other known optical beams, but from a practical point of view it was shown \cite{Dumke75}- \cite{Naqwi90}
  that certain laser sources produce fields that shows fundamental variations with respect to
  the canonical Gaussian beam. As well-known Gaussian beam is a minimun uncertainty field i.e. it possesses the minimum achievable angular spreading once
  the spatial extension is fixed; for certain laser sources, e.g. double heterojunction (DH) $Ga_{1-x}Al_x As$ lasers,
  which produce highly diverging fields, a Gaussian description
  for the transverse fields fails: in this case it was shown that a Lorentzian distribution
  is a better approximation, as it take into account of the higher angular spreading, being equal the spatial
  extension \cite{Naqwi90}. The paper is organized as follows: in next section we introduce the $LB$ and we study
  their propagation under Fresnel or paraxial approximation. In particular we give the closed-form expression for these kind of fields
  on a generical $(x,y)$ plane. Furthermore, we will make use
  of a theory group approach to introduce another class of optical beams obtained by multlipying a Lorentz beams with a two-dimensional Gaussian envelope.

\section{Lorentz Beams and paraxial propagation}
Let us suppose we have on a source plane, that we decide to be $(x,y,z=0)$ plane, the following scalar field distribution
\begin{equation}\label{LB z=0}
V_0(x,y)=\frac{A}{w_x w_y}\frac{1}{[1+(x/w_x)^2]}\frac{1}{[1+(y/w_y)^2]}
\end{equation}
where $A$ is a constant value and $w_x$ and $w_y$ are parameters related to
the beam width, with $A,w_x$ and $w_y \in \Re$. This kind of field is the
product of two functions of $x$ and $y$ variables which have the form of a Lorentzian function of parameter
$w_x$ and $w_y$. The Lorentzian is a well known bell-shaped curve used principally to describe
 the spectral lines of simple dinamical systems usually present in physics.
 Starting from the field in (\ref{LB z=0}) we wish to derive the form that such a field
 acquires during free propagation. To do this, we pass to the Fourier domain, calculating
 the plane waves spectrum on $z=0$.
 We have
\begin{eqnarray}\label{spettro z=0}
A_0(p,q) & =&\int_\infty\int_\infty[\frac{A}{w_x w_y}\frac{1}{[1+(x/w_x)^2]}\frac{1}{[1+(y/w_y)^2]} \exp(-2i\pi p x) \times \\
 & \times & \exp(-2\imath\pi q y) dx dy] \nonumber \\
& = &\frac{A}{w_x w_y}\int_\infty\frac{1}{[1+(x/w_x)^2]}\exp(-2i\pi p x)dx \times \\ & \times & \int_\infty\frac{1}{[1+(y/w_y)^2]}\exp(-2i\pi q y)dy \nonumber \\
& = & A\pi^2 \exp(-2\pi|p|w_x)\exp(-2\pi|q|w_y)
\end{eqnarray}
where $A_0(p,q)$ is the complex spectrum, $p$ and $q$ are the conjugated variables to $x$ and $y$ variables respectively.
Once we know the spectrum of plane waves on $z=0$ we can calculate it on a generical $(x,y)$-plane as follows
\begin{equation}\label{spettro z}
A_z(p,q)=A_0(p,q)\exp(2i\pi mz)=A\pi^2 \exp(-2\pi|p|w_x)\exp(-2\pi|q|w_y)\exp(2i\pi mz)
\end{equation}
because it is well known in which fashion a plane wave propagates in free-space. The parameter $m$ is a complex variable that
must fulfil the relationship
\begin{equation}\label{separazione}
m=\sqrt{(1/\lambda^2-p^2-q^2)}
\end{equation}
 and it is imaginary for evanescent waves, in which $p^2+q^2\geq 1/\lambda^2$, and real for homogeneous waves where
$p^2+q^2\leq 1/\lambda^2$, where $\lambda$ is the wavelength. We know that a field is said to be homogeneous when the spectrum $A_z(p,q)$ is
different from zero only inside
 the circle $p^2+q^2\leq1/\lambda^2$.
 \subsection{Fresnel or paraxial limit}
 Observing the (\ref{spettro z=0}) we deduce that if the values of $w_x$ and $w_y$ are enoughly greater than
 wavelength $\lambda$, the majority contribution to the field arises from homogeneous waves having the
 amplitude $A_z(p,q)$ corrisponding to points, in Fourier plane $(p,q)$,
 near to the origin. In this case one is authorized to do the following expansion
\begin{equation}\label{parax apprx}
m=\sqrt{(\frac{1}{\lambda^2}-p^2-q^2)}\simeq\frac{1}{\lambda}-\frac{(p^2+q^2)\lambda}{2}
\end{equation}
by retaining only the first two terms in the series,
so that the plane wave spectrum in (\ref{spettro z}) becomes
\begin{equation}\label{spettro parax}
A_z^{parax}(p,q)=A\pi^2 \exp(-2\pi|p|w_x-i\pi \lambda z p^2)\exp(-2\pi|q|w_y-i\pi \lambda z q^2)\exp(ikz)
\end{equation}
in which $k=2\pi/\lambda$ is the wave number.
Now, as we know the plane wave spectrum on $z$ we can also derive the full form of the field if we are able to inverte the two-dimensional
Fourier transform
\begin{eqnarray}\label{campo z}
V(x,y,z) & = & \int_\infty\int_\infty [A\pi^2\exp(ikz)\exp(-2\pi|p|w_x-i\pi \lambda z p^2) \times \nonumber \\
& \times & \exp(-2\pi|q|w_y-i\pi \lambda z q^2)]dpdq \nonumber \\
& = & A\pi^2\exp(ikz)\int_\infty\exp(-2\pi|p|w_x-i\pi \lambda z p^2)\exp(i2\pi p) dp \times \nonumber \\
& \times & \int_\infty\exp(-2\pi|q|w_y-i\pi \lambda z q^2)\exp(i2\pi q)dq
\end{eqnarray}
It is important to note that the role of paraxial approximation was to give a plane wave spectrum $A_z^{parax}(p,q)$
factorized in two terms, each one depending only from a single Fourier variable, $p$ or $q$; this is a properties that
 was not fulfilled by the exact spectrum in (\ref{spettro z}) for the presence of the term $\exp(i2\pi mz)$. As
 a consequence of that also the complex field $V(x,y,z)$ is in a similar factorized form.
To obtain a solution of (\ref{campo z}) let us focus on the integral
\begin{eqnarray}\label{Integrale 1}
I & = & \int_\infty\exp(-2\pi|p|w_x-i\pi \lambda z p^2)\exp(i2\pi p)dp
\end{eqnarray}
It can be written as
\begin{eqnarray}\label{Integrale 1}
I & = & \int_0^{\infty}\exp(-2\pi p w_x-i\pi \lambda z p^2)\exp(i2\pi p x)dp+\\
& + & \int_{-\infty}^0\exp(2\pi p w_x-i\pi \lambda z p^2)\exp(i2\pi p x)dp
\end{eqnarray}
For the first of two integral in right-hand side we have
\begin{eqnarray}\label{Integrale 2}
I & = & \int_0^{\infty}\exp(-2\pi p w_x-i\pi \lambda z p^2)\exp(i2\pi p x)dp= \nonumber \\
& = & \exp(c)\int_0^\infty\exp[-(ap+b)^2]dp=\frac{\exp(c)}{a}\int_b^\infty\exp(-s^2)ds \nonumber \\
& = & \frac{\exp(c)}{a}[\int_0^\infty\exp(-s^2)ds-\int_0^b\exp(-s^2) ds] \nonumber \\
& = & \frac{\sqrt\pi} {2}\frac{\exp[\pi(w_x-ix)^2/i \lambda z]}{\sqrt{i \pi \lambda z}}\{1-erf[\pi(w_x-ix)/\sqrt{i \pi \lambda z}]\}
\end{eqnarray}
where we have defined the following auxiliaries variables
\begin{eqnarray}\label{aux}
a^2=i\pi \lambda z  \\
b=\pi (w_x-ix)/a  \\
c=b^2 \\
s=ap+b
\end{eqnarray}
and $erf(x)$ is the usual error function $erf(x)=
(2/\sqrt{\pi})\int_0^x\exp(-s^2)ds$ \cite{Abramowitz65}. The other integral appearing
in (\ref{Integrale 1}) can be easily calculated by observing that
it is equal to that just derived in (\ref{Integrale 2}) after
having substituted the variable $x$ with $-x$. On utilizing this
result we finally obtain the full form of the field
\begin{eqnarray}\label{campo totale}
V(x,y,z) & = & \frac{A\pi^2}{4}\frac{\exp(ikz)}{i\lambda z}[V_x^+(x,z)+V_x^-(x,z)][V_y^+(y,z)+V_y^-(y,z)]
\end{eqnarray}
where
\begin{equation}\label{def V}
V_r^\pm(r,z)= \frac{\exp[\pi(w_r\pm ir)^2/i \lambda z]}{\sqrt{i \pi \lambda z}}\{1-erf[\pi(w_r\pm ir)/\sqrt{i \pi \lambda z}]\}
\end{equation}
and $r=x,y$.
Equation (\ref{campo totale}) is the principal result of the present work and
in next sections we analyse more in detail the propagation features of these beams.

\subsection{Propagation and diffraction-free range}
\begin{figure}[t1]
  \centerline{\includegraphics{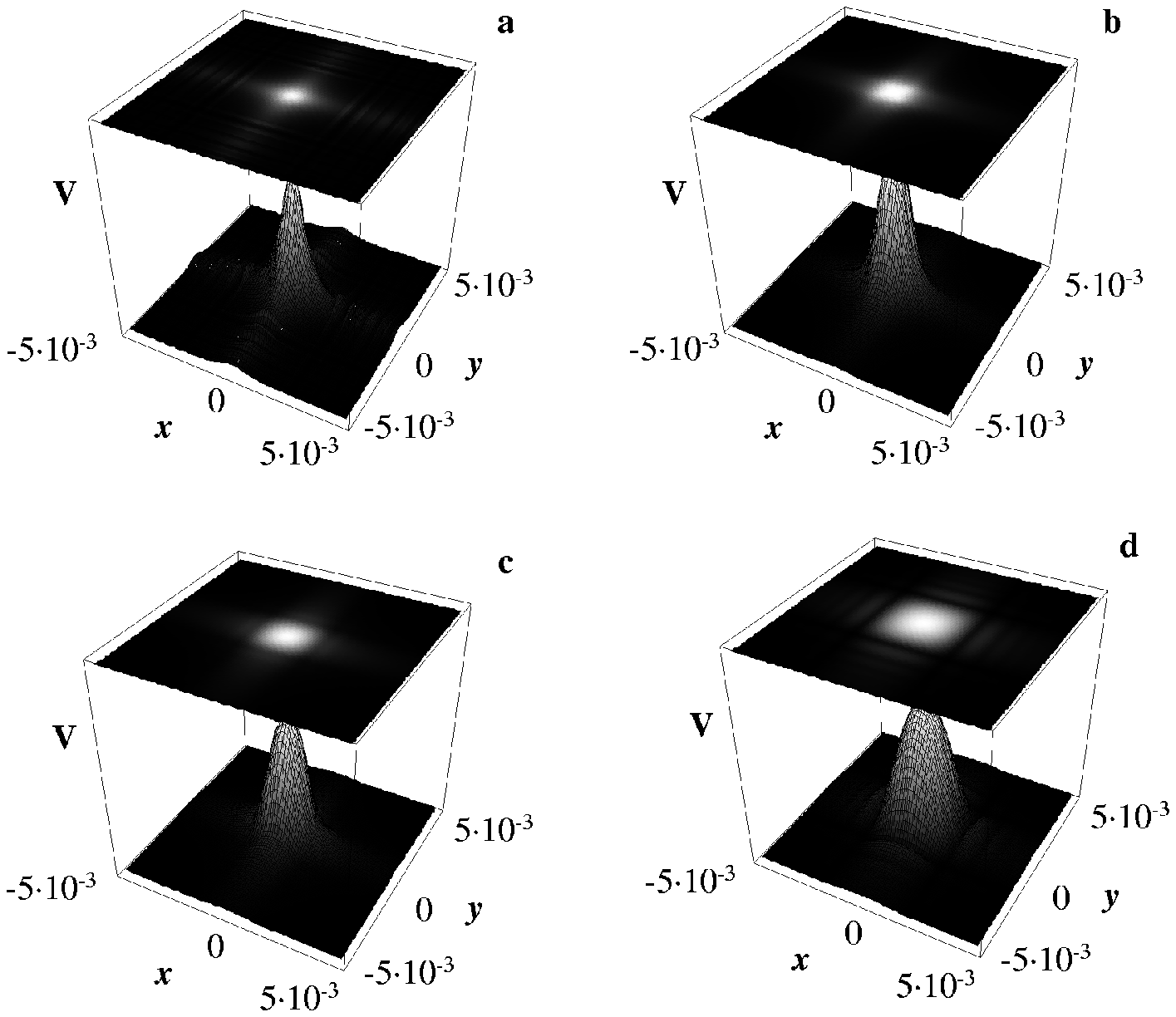}}
  \caption{Modulus of field's amplitude of a Lorentz beam on: a)source plane, i.e. $z=0$ b) when $z=0.1z_R$
  c) when $z=0.5z_R$ d) when $z=z_R$. The parameters are choosen as follows: $\lambda=0.6328 \mu m$,
  $w_x=w_y=w=10^3\lambda$. $z_R=2\pi w^2/\lambda=3.976 m$}
  \label{Modulo1}
\end{figure}

We expect that the field in (\ref{campo totale}) changes its shape during propagation
as a consequence of diffraction. It is well known, however, that it is possible to define a diffraction-free range, i.e. a linear distance
on $z$-axis, under which the beam remains essentially unchanged.
To do this we write the beam to values near to the source plane ($z=0$) by taking advantage of the expansion of error function
 for high values of its argument
\begin{equation}\label{erf}
erf(s)\approx1+\frac{s}{\pi}\exp(-s^2)\sum_{k=1}^\infty (-1)^k\frac{\Gamma(k-1/2)}{s^{2k}}
\end{equation}
with $|s|>>1$, $-\pi/2<\arg(s)<\pi/2$ and $\Gamma$ is the gamma special function \cite{Abramowitz65}.
On utilizing this expansion it is easy to verify that (\ref{campo totale}) reduces to (\ref{LB z=0}).
Indeed one obtains, by keeping only the first term
\begin{eqnarray}\label{range}
V(x,y,0) & =& \frac{A\pi^2}{4}\lim_{z\rightarrow 0 }\frac{1}{i\lambda z}[V_x^+(x,z)+V_x^-(x,z)][V_y^+(y,z)+V_y^-(y,z)]= \nonumber \\
& =& \frac{A\pi^2}{4}\lim_{z\rightarrow 0}\frac{1}{i\lambda z}[\frac{\Gamma(1/2)\sqrt{i\lambda z}}{\sqrt\pi\pi(w_x-ix)}+\frac{\Gamma(1/2)\sqrt{i\lambda z}}{\sqrt\pi\pi(w_x+ix)}] \times \nonumber \\
& &\times [\frac{\Gamma(1/2)\sqrt{i\lambda z}}{\sqrt\pi\pi(w_y-iy)}+\frac{\Gamma(1/2)\sqrt{i\lambda z}}{\sqrt\pi\pi(w_y+iy)}]= \nonumber \\
& = &\frac{A\pi^2}{4}[\frac{1}{w_x-ix}+\frac{1}{w_x+ix}][\frac{1}{w_y-iy}+\frac{1}{w_y+iy}]= \nonumber \\
& = &\frac{A}{w_x w_y}\frac{1}{[1+(x/w_x)^2]}\frac{1}{[1+(y/w_y)^2]}
\end{eqnarray}
where we used the identity $\Gamma(1/2)=\sqrt\pi$. The shape of
the field will be practically unchanged as far as the second term
in expansion will be negligible respect to the first. This
conducts us to the following condition,
\begin{equation}\label{condition1}
\frac{\Gamma(1/2)}{|\pi(w_r\pm ir)/\sqrt{i\pi\lambda z}|^2}>>\frac{\Gamma(3/2)}{|\pi(w_r\pm ir)/\sqrt{i\pi\lambda z}|^4}
\end{equation}
where $r=x,y$. Equation (\ref{condition1}) leads to
\begin{equation}\label{condition2}
\Gamma(1/2)>>\Gamma(3/2)\frac{\lambda}{\pi w^2}z
\end{equation}
and finally  (recall that $\Gamma(3/2)=\sqrt\pi/2$) we arrive  to the evaluation of the diffraction-free range (or Raileigh distance) for this kind of beams
\begin{equation}\label{zr}
z_R=\frac{2\pi w^2}{\lambda}
\end{equation}
In (\ref{condition2}) we have let $w_x=w_y=w$ to simplify the analysis. If one does not make this assumption
there will exist two different diffraction-free ranges, one to $x$-axis and another to $y$-axis.
In figure (\ref{Modulo1}) we report the modulus of field's amplitude
evaluated to different distances from the source.
In particular we observe that, for distances sufficiently shorter than $z_R$ the effect of diffraction is
neagligible, as expected, while when $z\approx z_R$ the diffraction affects the field shape.

\section{Lorentz-Gauss beams}
So far we have evaluated the paraxial propagation of the beam essentially by using an integral approach. In fact,
once we knew the field on the source plane $V_0(x,y)$ we were able to obtain it on every plane as dictated by Fresnel
theory, namely we had
\begin{equation}\label{fresnel}
V(x,y,z)=-\frac{i\exp(ikz)}{\lambda z}\int\int_\infty V_0(\xi,\eta)\exp[(i\frac{k}{2z}(x-\xi)^2+(y-\eta)^2)]d\xi d\eta
\end{equation}
Actually we used (\ref{fresnel}) in Fourier space to obtain the plane wave spectrum on $z$ and from it, by mean an
inverse Fourier transformation, we pointed out the field $V(x,y,z)$.
If we let
\begin{equation}\label{prodotto}
V(x,y,z)=\exp(ikz)f(x,y,z)
\end{equation}
where $f(x,y,z)$ is the slowly varying part of $V(x,y,z)$, we know that, if $V(x,y,z)$ fulfils the integral (\ref{fresnel}), then $f(x,y,z)$ fulfils a differential equation, known as
paraxial wave equation, that in tridimensional space holds
\begin{equation}\label{paraxial wave1}
\nabla_T^2f(x,y,z)+2ikf_z(x,y,z)=0
\end{equation}
where $\nabla_T^2$ is the two-dimensional Laplace operator in the transverse plane, i.e. $\nabla_T^2f=f_{xx}+f_{yy}$ and $f_z$ is the partial derivative respect to $z$ variable. This equation has very special properties,
and in particular we here are interested to its symmetry features. In fact there exists a beautiful theory,
due to the mathematician Sophus Lie, that allows to perform an analysis on the symmetry
groups associated to a particular differential equation (or, more in general, to a system of differential
equations). Essentially, the theory says that there exist some differential transformations which act
like operators on system's solutions and that lead to others solutions of the same equation,
when applied to an existing and known starting solution. The theory bases itself upon a certain number
of theorems in the context of differential geometry, and we will not enter in details;
however, in Appendix we put a proof of the derivation of the symmetry group we utilize in the following (see, for example, in \cite{Olver93}
for the theory foundations).
This theory
was used, for example, by Wunche \cite{Wunsche89} to show that Hermite-Gauss and Laguerre-Gauss beams (with complex argument) can be generated from
the fundamental Gaussian beam simply by applying to it the powers of certain differential Lie operators.
Among all the symmetry tranformations associated to (\ref{paraxial wave1}) we concentrate upon only one
which states that if $f(x,y,z)$ is a solution of the aforesaid equation also will be the function $f^*(x,y,z)$ where
\begin{equation}\label{lie}
f^*(x,y,z)=\frac{1}{1+iz/L}\exp[{-\frac{x^2+y^2}{w_0^2(1+iz/L)}}]f(\frac{x}{1+iz/L},\frac{y}{1+iz/L},\frac{z}{1+iz/L})
\end{equation}
with $w_0$ and $L$ two real parameter.
The property in (\ref{lie}) was recently proved, utilizing two different approaches, by Kiselev \cite{Kiselev04} that has utilized a
separation variable method and by
Gutierrez-Vega {\it et al.} \cite{Vega05}), in which the authors obtained the same result by inserting a
 well-constructed trial function into paraxial wave equation. It is interesting to note that the way we indicated here is a third different method, a method
 which possesses the merit to underline which is the foundation of this result, i.e. an hidden symmetry \cite{Wunsche89} and which
shows that all beams with a Gaussian envelope are intimately connected to the paraxial wave equation. As a consequence
 of (\ref{lie}) we immediately conclude that also exist the Lorentz-Gauss beams, namely
\begin{eqnarray}\label{Lo Gauss}
V^*(x,y,z) & = & \frac{A\pi^2}{4}\frac{\exp(ikz)}{i\lambda z}[V_x^+(\frac{x}{1+iz/L} ,\frac{z}{1+iz/L})+V_x^-(\frac{x}{1+iz/L},\frac{z}{1+iz/L})]\times \nonumber \\
& \times & [V_y^+(\frac{y}{1+iz/L},\frac{z}{1+iz/L})+V_y^-(\frac{y}{1+iz/L},\frac{z}{1+iz/L})] \times \nonumber \\
& \times & \exp[-\frac{x^2+y^2}{w_0^2(1+iz/L)}]
\end{eqnarray}
The shape of this kind of field depends by the choices of the
parameters $w_x,w_y,w_0$ and $L$. In particular if we choice
$w_0<w_x,w_y$, we obtain a beams that behaves like a Lorentz beam
near to the $z$ axis and like a Gaussian beam far from it. If we
put $L=z_R$, Lorentz-Gauss beams (\ref{Lo Gauss}) for $w_0=w_x
(w_y)$ shows a field shape more defined around $z$ axis with
respect to the Lorentz beams (\ref{campo totale}), i.e. on the
$x-y$ plane the LB has a wider bell-shaped field function with
respect to (\ref{Lo Gauss}).

\section{Conclusion}
A new kind of tridimensional, rectangularly-symmetric, realizable scalar
optical beams has been introduced. On the source plane
these beams are the product of two indipendent Lorentz function and
the exact analytical expression for the field on a generical plane, under
paraxial regime, has been derived. In particular it is interesting to note that
it deals of a rare case in which one is in the presence of physically realizable
fields, the propagation of which is known in closed-form,
that does not possess a Gaussian envelope term. This kind of field can represent
a valid candidate to modelize the shape of fields generated by certain
laser sources, as double heterojunction (DH) $Ga_{1-x}Al_x As$ lasers. Using a Lie group approach we
introduced also the solution obtained by modulating the aforesaid beams with a
Gaussian envelope.

\appendix\label{gruppi di Lie}

\section{Symmetry groups of scalar paraxial wave equation}
Consider the 3-D scalar paraxial wave equation
\begin{equation}\label{paraxial wave}
\nabla_T^2u(x,y,z)+2iku_z(x,y,z)=0
\end{equation}
We wish to show that this equation admits a class of solutions that are modulated by a two-dimensional Gaussian envelope. First of all,
to this equation it is possible to associate a manifold $M\in X\times U$, where $X=\Re ^p$, with $p=3$ meaning
number of independent variables and $U=\Re ^q$, with $q=1$ meaning the number of dependent variables.
On such a manifold it is also possible to define a tangent vector field in the following form
\begin{equation}\label{vector}
{\bf v}=\xi(x,y,z,u)\partial_x+\eta(x,y,z,u)\partial_y+\tau(x,y,z,u)\partial_z+\phi(x,y,z,u)\partial_u
\end{equation}
In the context of Lie group theory \cite{Olver93} tangent vector fields are the generators of symmetry transformations through the following relationship
\begin{equation}\label{psi}
\psi(\epsilon,{\bf v})=\exp(\epsilon {\bf v})
\end{equation}
where $\psi$ is the transformation induced by the vector ${\bf v}$ and $\epsilon$ is a real parameter characterizing the group.
A symmetry transformation is a map that allows to pass from a starting point $(x,y,z,u)$ on the manifold $M$ to another
point $(x',y',z',u')$ on the same manifold by mean the relation
\begin{equation}\label{map}
(x',y',z',u')= \psi(\epsilon,{\bf v}) (x,y,z,u)=\exp(\epsilon {\bf v}) (x,y,z,u)
\end{equation}
To find the explicit expression of the vector field in (\ref{vector}),
one can utilize the following procedure. It is necessary to extend the space $X \times U$
in order that it also contains the second order derivatives, an operation said {\it prolongation}. By consequence the prolongated tangent vector field becomes
\begin{eqnarray}\label{vector pro}
{\bf v} & = & \xi\partial_x+\eta\partial_y+\tau\partial_z+\phi\partial_u+\phi^x\partial_{u_x}+\phi^y\partial_{u_y}+\phi^z\partial_{u_z}+\phi^{xx}\partial_{u_{xx}}+ \nonumber \\
& + & \phi^{yy}\partial_{u_{yy}}+\phi^{zz}\partial_{u_{zz}}+\phi^{xy}\partial_{u_{xy}}+\phi^{xz}\partial_{u_{xz}}+\phi^{yz}\partial_{u_{yz}}
\end{eqnarray}
where we have dropped the dependence from the variables $x,y,z,u$.
We can rewrite it as
\begin{equation}\label{vector generico}
{\bf v}=\sum_{i=1}^p\xi_i\partial_{x_i}+\phi\partial_u\sum_J\phi^J\partial_{u_J}
\end{equation}
by defining $J=(j_1,j_2,...j_l)$, $1\leq j_l\leq p$ , $1\leq l \leq n$ where $n$ is the equation order.
All the coefficients in (\ref{vector pro}) are expressible in terms of $\xi, \eta, \tau$ and $\phi$ and their
derivatives through the formula
\begin{equation}\label{coeff_gen}
\phi^J=D_J(\phi-\sum_{i=1}^p\xi^iu_{J,i})+\sum_{i=1}^p\xi^i u_{J,i}
\end{equation}
where $u_i=\partial u/\partial x_i$ and $u_{J,i}=\partial u_J/\partial x_i$ with $x_i$ generical variable and $D_J$ representing the
total derivative.
Among all such 13 coefficients only $\phi^{xx},\phi^{yy},\phi^{z}$ are important for our purpose.
Indeed, under certain hypotheses, it is possible to obtain all the symmetry group of transformation of scalar paraxial wave equation through the
condition
\begin{equation}\label{cond}
{\bf v}[u_{xx}+u_{yy}+2iku_z]=0
\end{equation}
that, by taking into account relation (\ref{vector pro}) implies
\begin{equation}\label{cond 2}
\phi^{xx}+\phi^{yy}+2ik\phi^z=0
\end{equation}
On performing the calculations as in (\ref{coeff_gen}) we find
\begin{eqnarray}\label{phi_xx}
\phi^{xx}&=&\phi_{xx}+u_x(2\phi_{xx}-\xi_{xx})+u_y(-\eta_{xx})+u_z(-\tau_{xx})+u_x^2(\phi^{uu}-2\xi_{uu})+\nonumber \\
& + & u_x u_y(-2\eta_{xu})+u_x u_z(-2\tau_{xu})+u_{xx}u_x(-3\xi_u)+u_{xx}u_y(-\eta_u)+\nonumber \\
& +& u_{xx}u_z(-\tau_u)+u_{xy}u_x(-2\eta_u)+u_{xx}(\phi_u-2\xi_x)+u_x^3(-3\xi_{uu})+ \nonumber \\
& + & u_{yx}(-2\eta_x)+u_{zx}(-2\tau_x)+u_x u_{zx}(-2\tau_u)+u_x^2 u_y (-\eta_{uu})+ \nonumber \\
& + & u_x^2 u_z (-\tau_{uu})
\end{eqnarray}
\begin{eqnarray}\label{phi_yy}
\phi^{yy}&=&\phi_{yy}+u_y(2\phi_{yu}-\eta_{yy})+u_x(-\xi_{yy})+u_z(-\tau_{yy})+u_y^2(\phi^{uu}-2\eta_{yu})+\nonumber \\
& + & u_x u_y(-2\xi_{yu})+u_y u_z(-2\tau_{yu})+u_{yy}u_y(-3\eta_u)+u_{yy}u_x(-\xi_u)+\nonumber \\
& +& u_{yy}u_z(-\tau_u)+u_{xy}u_y(-2\xi_u)+u_{yy}(\phi_u-2\eta_y)+u_y^3(-3\eta_{uu})+ \nonumber \\
& + & u_{yx}(-2\xi_y)+u_{zy}(-2\tau_u)+u_y u_{zy}(-2\tau_u)+u_y^2 u_x (-\xi_{uu})+ \nonumber \\
& + & u_y^2 u_z (-\tau_{uu})
\end{eqnarray}
\begin{eqnarray}\label{phi_z}
\phi^{z}&=&\phi_{z}+u_z(2\phi_{u}-\tau_{z})+u_x(-\xi_{z})+u_zu_x(-\xi_u)+u_y(-\eta_z)+\nonumber \\
& + & u_z u_y(-\eta_u)+u_z^2(-\tau_u)
\end{eqnarray}
Now we equate the right and left-hand  side homologous terms appearing in (\ref{cond 2}) and finally we obtain

\begin{equation}\label{coeff 1}
\phi_{xx}+\phi_{yy}=\phi_t
\end{equation}
\begin{equation}\label{coeff 2}
2(\phi_{xu}-\xi_{xx})-\xi_{yy}=-\xi_t
\end{equation}
\begin{equation}\label{coeff 3}
2(\phi_{yu}-\xi_{xx})-\xi_{xx}=-\xi_y
\end{equation}
\begin{equation}\label{coeff 4}
2\xi_x=\tau_t
\end{equation}
\begin{equation}\label{coeff 5}
2\eta_y=\tau_t
\end{equation}
\begin{equation}\label{coeff 6}
\tau_x=\tau_y=\tau_u=0
\end{equation}
\begin{equation}\label{coeff 7}
\eta_u=\xi_u=0
\end{equation}
\begin{equation}\label{coeff 8}
\phi_{uu}=0
\end{equation}

where we have defined an auxiliary variable $t=z/(2ik)$.
Solving this system of equations is not difficult but we report here the result only, that can be verified by
substitution,
\begin{equation}\label{tau}
\tau=c_1+2c_2t+4c_3t^2
\end{equation}
\begin{equation}\label{xi}
\xi=c_2x+4c_3tx-2c_4t+c_5
\end{equation}
\begin{equation}\label{eta}
\eta=c_2y+4c_3ty-2c_4t+c_6
\end{equation}
\begin{equation}\label{phi}
\phi=[-c_3(x^2+y^2)+c_4(x+y)-4c_3t+c_7]u+\alpha(x,y,t)
\end{equation}
where $\alpha$ is a generical function and $c_i (i=1,2,...7)$ are integration constants.
Each of these constants is related to a particular generator of symmetry
${\bf v}_i$ which can be obtained by letting tidily all the constants to zero except the $i$th.
Among all such a transformations there is the following one
\begin{equation}\label{v}
{\bf v}=4tx\partial_x+4ty\partial_y+4t^2\partial_t+(-x^2-y^2-4t)u\partial_u
\end{equation}
From (\ref{map})-(\ref{v}) follows that, if we indicate as $u(x,y,s)$ a solution then also the following one
represents a valid one
\begin{equation}\label{sol1}
u^*(x,y,s)=\frac{1}{1+4\epsilon s}\exp[{-\epsilon\frac{x^2+y^2}{(1+4\epsilon s)}}]u(\frac{x}{1+4\epsilon s},\frac{y}{1+4 \epsilon s},\frac{z}{1+4\epsilon s})
\end{equation}
On coming back to the old coordinate z and letting $\epsilon/2k=1/L$ and $w_0^2=1/\epsilon$, it holds
\begin{equation}\label{sol}
u^*(x,y,z)=\frac{1}{1+iz/L}\exp[{-\frac{x^2+y^2}{w_0^2(1+iz/L)}}]u(\frac{x}{1+iz/L},\frac{y}{1+iz/L},\frac{z}{1+iz/L})
\end{equation}
that represents the result we were looking for.

\end{document}